\documentclass[aps,prc,twocolumn,superscriptaddress,showpacs,floatfix]{revtex4-1}

\usepackage{amssymb}
\usepackage{graphicx}
\usepackage{ulem}

\newcommand{\gsim}{\, \raisebox{-0.8ex}{$\stackrel{\textstyle >}{\sim}$ }}
\newcommand{\lsim}{\, \, \raisebox{-0.8ex}{$\stackrel{\textstyle <}{\sim}$ }}

\begin{document}
\title{The maximum mass and radius of neutron stars and the nuclear symmetry energy}

\author{S. Gandolfi}
\affiliation{Theoretical Division, Los Alamos National Laboratory, Los Alamos, New Mexico 87545, USA}

\author{J. Carlson}
\affiliation{Theoretical Division, Los Alamos National Laboratory, Los Alamos, New Mexico 87545, USA}

\author{Sanjay Reddy}

\affiliation{Theoretical Division, Los Alamos National Laboratory, Los Alamos, New Mexico 87545, USA}
\affiliation{Institute for Nuclear Theory, University of Washington, Seattle, Washington 98195-1550, USA}

\begin{abstract}
We calculate the equation of state of neutron matter with realistic two-
and three-nucleon interactions using quantum Monte Carlo techniques, and
illustrate that the short-range three-neutron interaction determines the
correlation between neutron matter energy at nuclear saturation density
and higher densities relevant to neutron stars.  Our model also makes
an experimentally testable prediction for the correlation between the
nuclear symmetry energy and its density dependence -- determined solely by
the strength of the short-range terms in the three neutron force. The same force provides
a significant constraint on the maximum mass and radius of neutron stars.
\end{abstract}


% 21.65.Cd 	Asymmetric matter, neutron matter 
% 21.65.Ef 	Symmetry energy 
% 26.60.-c 	Nuclear matter aspects of neutron stars
% 26.60.Kp      Equations of state of neutron-star matter 
% 21.65.-f 	Nuclear matter
% 21.65.Mn 	Equations of state of nuclear matter

\maketitle

Since their discovery, neutron stars have remained our sole laboratory to
study matter at supra-nuclear density and relatively low temperature. The
equation of state (EoS) of matter at these densities is largely unknown
but uniquely determines the structure of neutron stars and the relation
between their mass ($M$) and radius ($R$).  Matter that can support large
pressure for a given energy density (typically called a stiff EoS) will
favor large neutron star radii for a given mass. Such an EoS also predicts
large values for the maximum mass of a neutron star that is stable with respect
to gravitational collapse to a black hole. Conversely, a high density phase
that predicts a smaller pressure will result in more compact neutron
stars and smaller maximum masses.

The recent accurate measurement of a large neutron star mass $M=1.97
\pm 0.04 M_{\rm solar}$ in the system J1614-2230 provides
strong evidence that the high density equation of state is stiff
\cite{Ransom:2010}.  Interestingly, attempts to infer neutron star
radii have favored relatively small values ranging from 9 to 12 km
\cite{Webb:2007,Guver:2010,Steiner:2010}. Although the radius inference
depends on specific model assumptions, these smaller radii imply  a
soft EoS in the vicinity of nuclear saturation density.  Taken together,
they indicate that the EoS of dense matter makes a transition from soft
to stiff at supra-nuclear density. In this Rapid Communication we show that the
three-neutron force (3n) is the key microscopic ingredient that determines
the nature of this transition.

The importance of three-body forces in nuclear physics is well known, and
quantum Monte Carlo (QMC) calculations of light nuclei have clarified its structure and
strength. However, in these systems the dominant three-body force acts between 
two neutrons and proton or between two protons and a neutron. While the force among three neutrons 
is important in light neutron-rich nuclei, the short distance behavior is not easily accessible \cite{Pieper:2001}.  Properties of large
neutron-rich nuclei are potentially sensitive to this interaction,
especially if the symmetry energy provides a reliable measure of the
energy difference between pure neutron matter and symmetric nuclear
matter at saturation density.  There has been much recent progress
in both theory and experiments to measure the symmetry energy and its
density dependence, as reviewed in Refs.~\cite{Steiner:2005,Shetty:2007}.
The symmetry energy is expected to be in the range $32\pm2$ MeV.
We explore this experimentally suggested range for the nuclear symmetry
energy and show that a more precise determination is needed to adequately
constrain the 3n interaction.

In this work we solve the non-perturbative many-body nuclear Hamiltonian
using the auxiliary field diffusion Monte Carlo (AFDMC)~\cite{Schmidt:1999}
method.  Its accuracy in studying nuclear systems has been tested in
light nuclei \cite{Gandolfi:2007}.
The extension to include three-body forces in pure neutron rich 
systems is straightforward with no additional approximations within the AFDMC 
technique~\cite{Sarsa:2003}, and a comparison with the Green's
function Monte Carlo (GFMC) has been extensively tested in neutron
drops~\cite{Gandolfi:2011}.
We present results for the EoS of
neutron matter using phenomenological two-neutron (2n) potentials, which
provide an accurate description of nucleon-nucleon scattering data
up to high energies, and study the role of the poorly constrained 3n
interaction.  

In earlier work it has been established that the EoS in
the density regime $(1-3)\rho_0$ plays an essential role in determining
the neutron star radius \cite{Lattimer:2001}. In this density regime,
the 3n interaction plays a critical role because of a large cancellation
between the attractive and repulsive parts of the 2n interaction arising
from the long and short distance behavior, respectively. Consequently, we
find that the neutron star radius for a canonical mass of $1.4$ M$_{\rm
solar}$ is especially sensitive to the 3n interaction.  Although matter
in the neutron star will contain a small admixture of protons, here we
calculate the EoS of pure neutron matter for the following reasons. First,
the structure of the interactions between neutrons is simpler than those
between neutron and protons.  Second, these simpler interactions are
amenable to QMC methods to solve the many-body problem as it is devoid of
the complexities of the isospin dependent spin-orbit  and three-nucleon
potentials, and clustering effects likely in systems with protons.
Third, the fraction of protons required to ensure stability is small and
is typically less than $10$\%. Finally, since generically neutron matter has 
higher pressure than matter containing any fraction of protons or 
strangeness in the form of hyperons or kaons, our results provide stringent 
upper bounds on the neutron maximum mass and radius.

To compute the EoS for neutron stars it is necessary to describe the
nucleon-nucleon interactions at short distances or large relative 
momenta up to $p \simeq 2 p_{Fn} \simeq 660~{\rm MeV} (\rho/\rho_0)^{1/3}$,
where $p_{Fn}$ is the Fermi momentum, $\rho$ is the
typical density in the neutron star core, and $\rho_0=0.16$ fm$^{-3}$
is the nuclear saturation density.
Relative momenta up to $p_{Fn}$ are required in even a mean-field (Fermi gas)
description, and the nn interaction scatters nucleons to larger momenta 
up to order (1.5-2)$p_{Fn}$ at saturation density.  Descriptions of higher density neutron matter with
softer interactions if they are consistently evolved to lower scales,
must include 3n (and potentially 4n) interactions.

Phenomenological two nucleon
potentials such as the Argonne potential have been constructed to
describe scattering data up to relative momenta $\simeq 600$ MeV
with high accuracy \cite{Wiringa:1995}. 
Despite the fact that the Argonne potential has been fit 
up to laboratory energies of 350~{\rm MeV}, it very well reproduces scattering
data up to much larger energies~\cite{BobPrivate} The AV8' interaction
we employ in this study is identical to the full AV18 interaction in $s$ and 
$p$ waves, and includes the dominant one-pion interaction in higher partial
waves.  Chiral interactions also reproduce the scattering data
very well below 350~{\rm MeV} laboratory energy, but they fail rapidly above because of the
cutoff in presently available interactions.
At larger momentum transfer,
the potentials cannot describe inelasticities, but in scattering
channels where inelasticities are known to be small they 
have been shown to provide a good description.  They also provide good
predictions~\cite{Schiavilla:2007} of high-momentum components of nuclear
wave functions as observed in nucleon \cite{Piasetzky:2006,Tang:2003} and
electron scattering\cite{Subedi:2008,Shneor:2007}.  These high momentum
observables provide a test of the assumed short-distance features. In
the low-energy high-momentum region relevant to neutron stars the
inelasticities in 2n scattering must be absorbed into many-body forces
(3n, 4n, $\dots$) intimately connected to the short-distance behavior of
the 2n interaction.

The nuclear Hamiltonians we consider contain the non relativistic kinetic energy,
and the 2n and 3n interactions:
\begin{equation}
H=-\frac{\nabla^2}{2m} + V_{2n} + V_{3n} \,.
\end{equation}
For the 2n potential, we use the Argonne AV8' model~\cite{Wiringa:2002}
and the form of the 3n interaction is inspired by both the Urbana IX
and the  Illinois models \cite{Pieper:2001}. We consider a range of 3n
interactions that contain long-distance $s$ and $p$ wave $2\pi$ exchange
contributions,
an intermediate-range ($3\pi$ loops)
contribution, and a spin-independent short-range repulsive term.
Explicitly,
\begin{equation}
V_{3N} = A^{PW}_{2\pi} {\cal O}^{2\pi,PW}+ A^{SW}_{2\pi} 
{\cal O}^{2\pi,SW}+ A_{3\pi} {\cal O}^{3\pi} + A_{R} {\cal O}^{R} \,.
\end{equation}
This form of interaction includes all the terms present in low
order chiral interaction, plus selected terms found to be important
in studies of light nuclei and nuclear matter using the Argonne
interactions.

The structure of the operators ${\cal O}$ appearing above are defined
in Ref.~\cite{Pieper:2001}. The relative contributions of these four
components of the 3n force depends on the 2n interaction.  We find
that for the Argonne potential, the 2n interactions suppress the
long-distance ($2\pi$) contribution of the 3n force in the ground state. This
suppression is a result of the pion-range correlations induced by
the 2n force, we find it also occurs for the super-soft core $NN$ 
interaction \cite{Pieper}. For typical ranges of values of the strength parameters $ A^{PW}_{2\pi}$
and $A^{SW}_{2\pi}$ considered in Ref.~\cite{Pieper:2001} we find
the contribution of these operators to the ground state energy is
repulsive but very small at all densities studied. In contrast, this
interaction is large and attractive in light nuclei where both neutrons
and protons contribute.  The intermediate-range ($3\pi$) 3n interaction was
introduced to fit the properties of weakly bound neutron-rich nuclei such as
$^8$He~\cite{Pieper:2001}. Earlier calculations \cite{Sarsa:2003} have
shown that this interaction is strong and attractive in neutron matter
for typical values of $ A_{3\pi}$ quoted in Ref. \cite{Pieper:2001}. In
this work, we explored a range of values for $ A_{3\pi}$ from
zero to that in the Illinois-7 3n interaction~\cite{Pieper:2008} because
the structure of this term is still not fully understood or constrained.
We use a phenomenological short-range repulsive term as in the Urbana and
Illinois three-body forces, with $V_R=A_R{\cal O}^R=A_R\sum_{cyc}T^2(m_\pi
r_{ij})T^2(m_\pi r_{jk})$, where the function $T(x)$ is defined in
Ref.~\cite{Pieper:2001}.  We have also considered a different form
$V_{\mu}^R=A_R\sum_{cyc}v(r_{ij})v(r_{jk})$ with and $v(r)=\exp(-2\mu
r)$; other different forms of $V_R$ have been explored, giving very
similar results.  

The 3n interaction we employ is not intended to be a microscopic
treatment of the complete 3n interaction.  It assumes that for the neutron
matter equation of state the effects of  more complicated spin-dependent 
short-distance 3n interactions, relativistic effects, 
and potential 4n interactions can be mimicked  with simplified three-neutron 
interactions with a wide range of spatial dependence. This assumption
has been tested in the case of relativistic corrections, where in Ref.~\cite{Akmal:1998}
it was found that the density dependence of the relativistic effects is similar
to that of the 3n interaction. Further tests of the density dependence of
specific higher-order terms in the chiral interaction are valuable.
The different forms of $V_R$ we have explored span a wide range
of density dependence for the 3n interaction, as shown below.

For the 3n interaction we vary both $ A_{3\pi}$ and
$\mu$ to study the sensitivity to short-range physics.  The strength of
the short-range 3n interaction  $A_{R}$ is taken to be a free parameter
adjusted to yield the  experimentally accessible nuclear symmetry
energy.  Although not proven, we make the following reasonable assumptions: (1)
relativistic effects in neutron matter show a similar density dependence
to the short-range three-nucleon interaction as carefully studied in
Ref. ~\cite{Akmal:1998}, (2) the density dependence of additional 
spin-dependent short-range 3n interactions (for example, higher-order 
terms in chiral expansions) in the equation of state of neutron matter
can be described in a spin-independent model,  
and (3) four-nucleon force contributions with different density dependence are
suppressed relative to the 3n force for densities up to $(2-3)\rho_0$.
This last assumption can be justified at nuclear density by the high-precision
fits to light-nuclei obtained with only 3n forces~\cite{Epelbaum:2009};
at higher density this model assumption can be tested by its predicted
correlation between properties of neutron-rich nuclei and neutron stars.

\begin{figure}[t]
\centering
\includegraphics[width=0.45\textwidth]{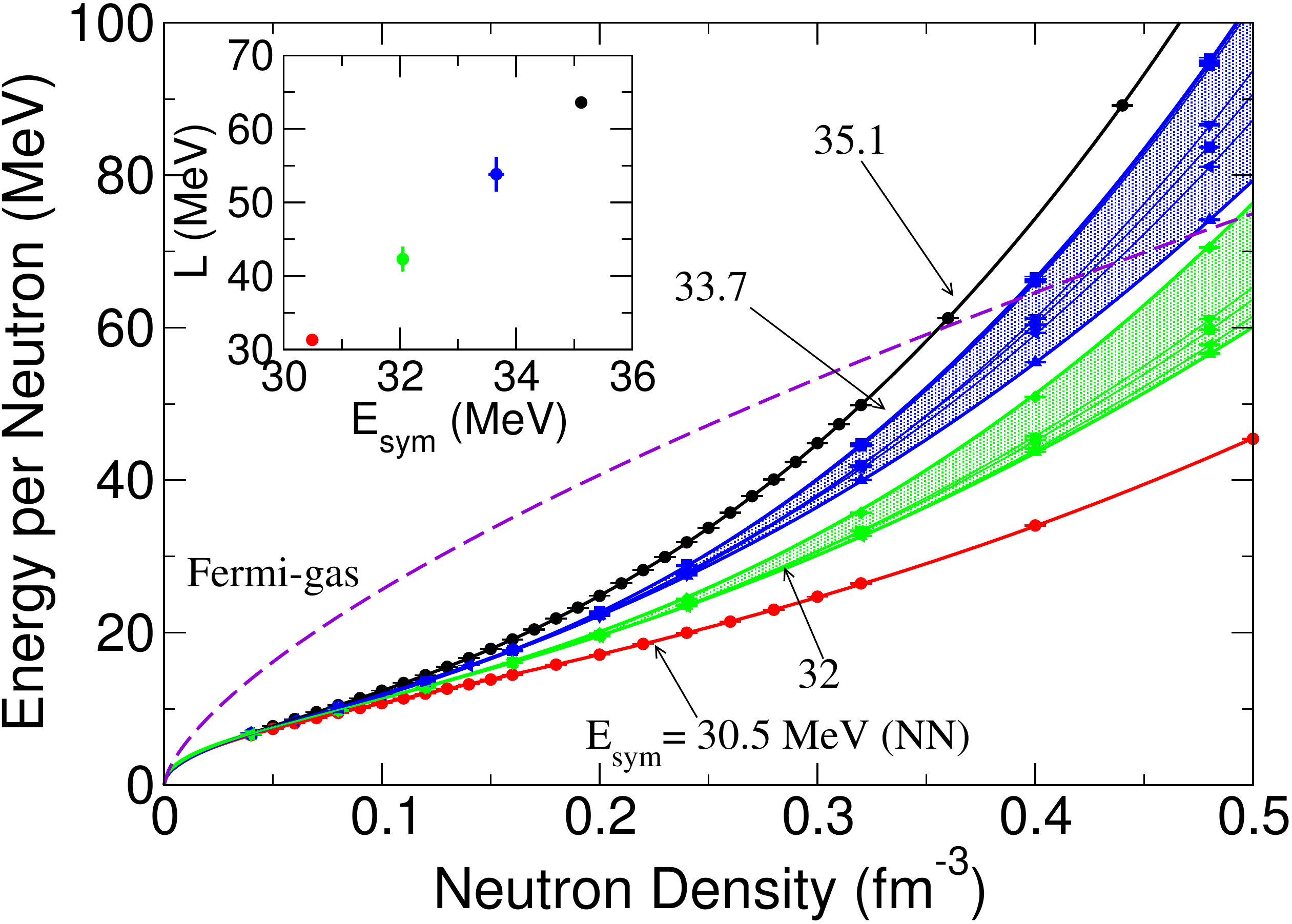} 
\caption{(Color online) The energy per particle of neutron matter for different values 
of the nuclear symmetry energy ($E_{\rm sym}$). 
For each value of $E_{\rm sym}$ the corresponding band shows the effect of
different spatial and spin structures of the three-neutron interaction.
The inset shows the linear correlation between $E_{\rm sym}$ and its 
density derivative $L$.
}  
\label{fig:eos}
\end{figure}
We assume that $E_{\rm sym}=E_{\rm neutron}(\rho_0) -E_{\rm
nuclear}(\rho_0)$ and using experimental values of $E_{\rm sym}=32 \pm
2$ MeV ~\cite{Tsang:2009} and $E_{\rm nuclear}(\rho_0)=-16.0 \pm 0.1$
MeV from nuclear masses models \cite{Moller:1995} we obtain an empirical
constraint for neutron matter energy $E_{\rm neutron}(\rho_0) = 16 \pm2$
MeV.  Potential higher-order corrections to the quadratic nuclear symmetry 
energy, for which there is some theoretical motivation but no clear experimental evidence,  
may affect the extraction of the neutron matter energy and increase the associated error. 
In this work we ignore these poorly known corrections and tune $A_{R}$ to reproduce the 
neutron matter energy in the range $ 16 \pm2$ MeV.   Our results are shown in Fig.~\ref{fig:eos}, where the green and
blue points are QMC results for different choices of $A_R$ corresponding
to $E_{\rm neutron}(\rho_0)=16$ MeV ( $E_{\rm sym}=32$ MeV) and $E_{\rm
neutron}(\rho_0)=17.7$ MeV ( $E_{\rm sym}=33.7$ MeV), respectively.
The results are compared to those obtained using a 2n force without 3n
($E_{\rm sym}=30.5$ MeV), and 2n combined with the Urbana IX 3n ($E_{\rm
sym}=35.1$ MeV). The bands depict the sensitivity to short-distance spin
and spatial structure of the 3n interaction and are obtained by varying
the range of the 3n short-distance force and $A_{3\pi}$.

In the vicinity of nuclear density, $E_{\rm neutron}(\rho)= E_{\rm
neutron}(\rho_0)+L/3~(\rho-\rho_0)/\rho_0$ where $L$ is related to the
derivative of the nuclear symmetry energy. The inset in Fig.~\ref{fig:eos}
shows the correlation between $E_{\rm sym}$ and $L$. This 
correlation is insensitive to the large variations in the range of the
short-range 3n force $\mu$ and the strength of the $3\pi$ term  $
A_{3\pi}$.  This is in sharp contrast to the predictions of mean field theories where 
the slope was found to be very sensitive to the choice of effective interactions \cite{Brown:2000}. 
Previous calculations of neutron matter up to $\rho_0$\cite{Hebeler:2010}
use a chiral 2n interaction fit to laboratory energies of 350 MeV plus the two-pion
exchange three-nucleon interaction to calculate the neutron matter equation
of state using perturbation theory.  In contrast to our results, a significant repulsion from
the $2\pi$ exchange long-range 3n interaction was found. 
Since this force is  better constrained by light nuclei, these earlier calculations can make a prediction for the 
neutron matter energy independent of the phenomenological short-range interaction, which plays an important 
role in our calculation. To understand this basic difference, further tests 
of the convergence of perturbation theory and the chiral expansion 
in the diagrammatic calculations, a survey of other two-body interactions 
in the AFDMC, and the incorporation of chiral interactions in non-perturbative methods
such as lattice and suitable extension of QMC would be necessary.

Current determinations of $L$ have relied on analysis of neutron-skins, 
surface contributions to the symmetry energy of neutron-rich nuclei,
and isospin diffusion in heavy-ion reactions.  These studies have been
useful, but not very constraining as acceptable values are in the range
$L=40-100$ MeV~\cite{Tsang:2009}. However, a better determination of $L$ even 
with modest reduction in the error would test our model for 2n and 3n interactions.

The predictions
of QMC can be accurately fit using   
\begin{equation}
E(\rho) = a~\left(\frac{\rho}{\rho_0}\right)^\alpha
~+~b~\left(\frac{\rho}{\rho_0}\right)^\beta \,, 
\label{eq:fit}
\end{equation} 
where the coefficients $a$ and $\alpha$ are sensitive to the low density
behavior of the EoS, while  $b$ and $\beta$ are sensitive to the high
density physics \cite{Gandolfi:2009}. We find that the 3n force plays
a key role in determining the coefficient $b$ and the variation of
the other EoS parameters is comparatively small. Numerical values for
these parameters are reported in Table \ref{tab:fitpara} for selected
Hamiltonians.

\begin{table}[htbp]
\centering
\begin{tabular}{@{} lcccccc @{}}
\hline
$3N$ force       & $E_{\rm sym}$ &$L$ & $a$    &  $\alpha$ & $b$ & $\beta$ \\
               & (MeV)  & (MeV) & (MeV)    &        & (MeV)  & \\
\hline
none           & 30.5   & 31.3  &   12.7   &  0.49  &  1.78  &  2.26 \\
$V_{2\pi}^{PW}+V^R_{\mu=150}$   & 32.1   & 40.8  &   12.7   &  0.48  &  3.45  &  2.12 \\
$V_{2\pi}^{PW}+V^R_{\mu=300}$   & 32.0   & 40.6  &   12.8   &  0.488 &  3.19  &  2.20 \\
$V_{3\pi}+V_R$ & 32.0   & 44.0  &   13.0   &  0.49  &  3.21  &  2.47 \\
$V_{2\pi}^{PW}+V^R_{\mu=150}$   & 33.7   & 51.5  &   12.6   &  0.475 &  5.16  &  2.12 \\
$V_{3\pi}+V_R$ & 33.8   & 56.2  &   13.0   &  0.50  &  4.71  &  2.49 \\
UIX            & 35.1   & 63.6  &   13.4   &  0.514 &  5.62  &  2.436 \\
\hline
\end{tabular}
\caption{Fitting parameters for the neutron matter EoS defined in Eq.~\ref{eq:fit} 
for selected different Hamiltonians.}
\label{tab:fitpara}
\end{table}

\begin{figure}[h]
\centering
\includegraphics[width=0.45\textwidth]{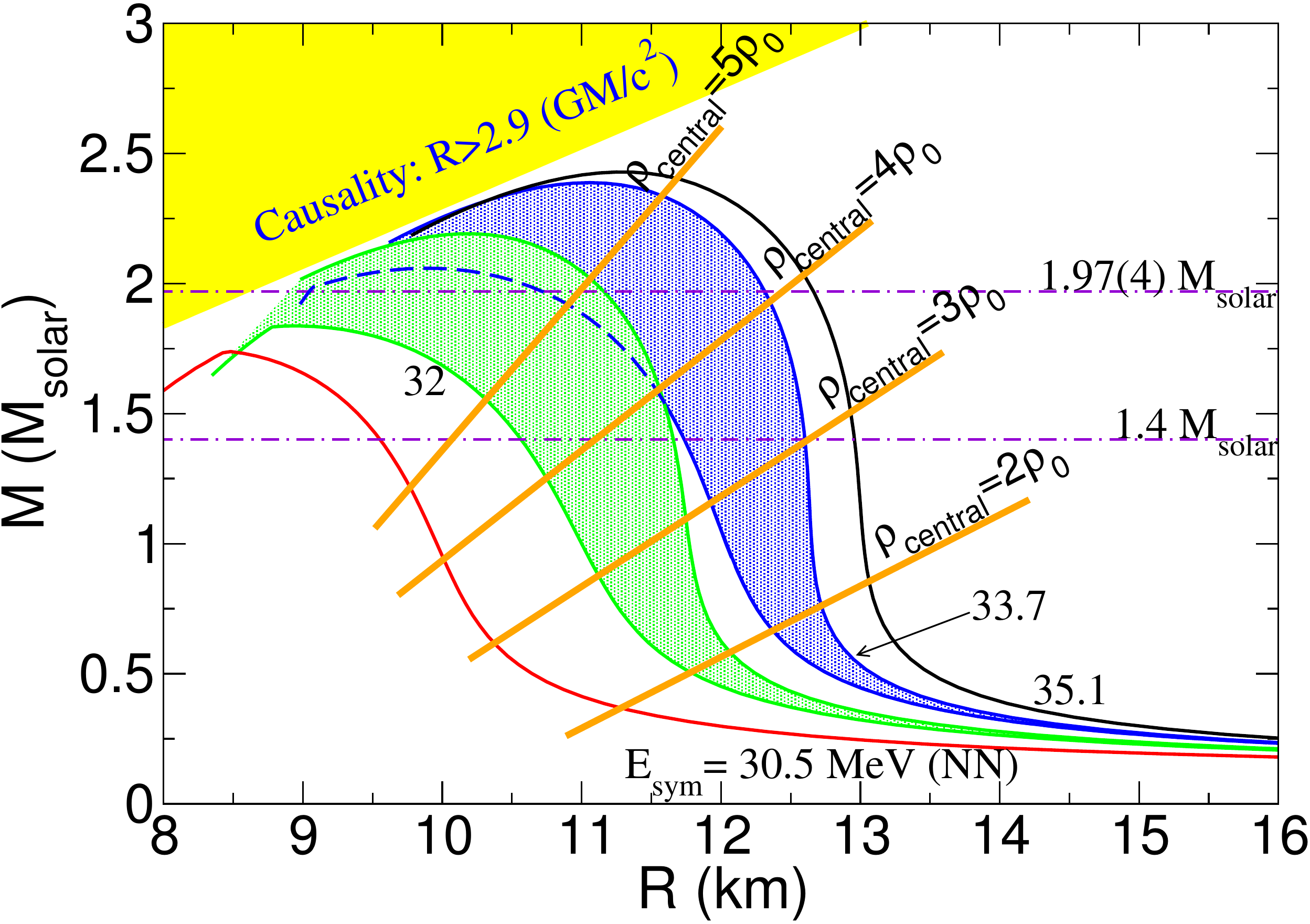} 
\caption{(Color online) Mass-radius relation for the EoS with three-neutron
interactions corresponding to the bands for different $E_{\rm sym}$
shown in Fig.~\ref{fig:eos}. The intersections with the orange lines
roughly indicate central densities realized in these stars.
}  
\label{fig:MR}
\end{figure} 
To calculate the mass and radius of neutron stars we solve the
Tolman-Oppenheimer-Volkoff (TOV) equations for the hydrostatic structure
of a spherical non rotating star using the QMC equation of state for
neutron matter \cite{Lattimer:2004,Gandolfi:2010}.  The QMC EoS we use
is for $\rho \ge \rho_{\rm crust}=0.08$ fm$^{-3}$. Below this density we use
the EoS of the crust obtained in earlier works in Refs. \cite{BPS:1971}
and \cite{NegeleVautherin:1973}.

The neutron star mass-radius predictions are obtained by varying the
3n force and are shown in Fig.~\ref{fig:MR}. The striking feature is the
estimated error in the neutron star radius with a canonical mass of
$1.4$ M$_{\rm solar}$. The uncertainty in the measured symmetry energy
of $\pm 2$ MeV leads to an uncertainty of about 3 km for the radius,
while the uncertainties in the short-distance structure of the 3n force
predicts  a radius uncertainty of $\lsim1$ km.  The different bands of
Fig. \ref{fig:MR} correspond to the EoS of Fig. \ref{fig:eos} with the same
colors, giving different values of $E_{\rm sym}$.

\begin{figure}[h]
\centering
\includegraphics[width=0.47\textwidth]{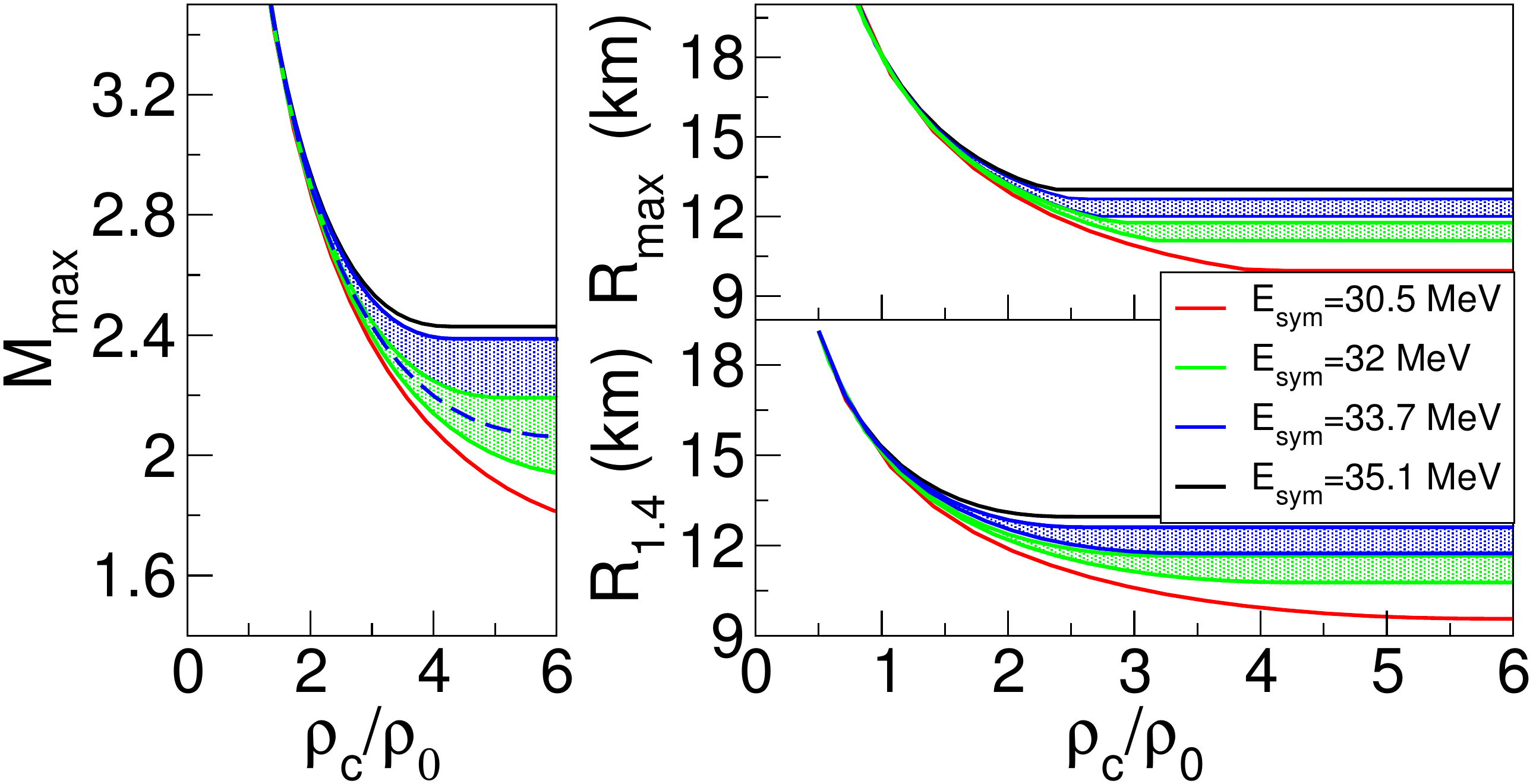} 
\caption{(Color online) Bounds on the maximum mass and radius for different equations
of state as a function of the critical
density $\rho_c$. The left panel shows the maximum mass; the right top
and bottom panels show the maximum possible radius for any neutron star
with mass greater than $1.2 M_{\rm solar}$ 
and for a neutron star with $M=1.4 M_{\rm solar}$, respectively. }
\label{fig:maxMR}
\end{figure} 

The central density of stars with $M\gsim 1.5 M_{\rm solar}$ are larger
than 3$\rho_0$. At these higher densities, effects such as
relativistic corrections to the kinetic energy, retardation 
in the potential, and four- and higher body forces become important.
Consequently, non-relativistic models violate causality and predict
a sound speed $c_s = \sqrt{\partial p/\partial \epsilon} \gsim c $
for $\rho \simeq (4-5) \rho_0$.  To overcome this deficiency we adopt
the strategy suggested in Ref.~\cite{RhoadesRuffini:1974} and replace
the EoS above a critical density  $\rho_c$ by the maximally stiff or
causal EoS given by $p(\epsilon)=c^2 \epsilon - \epsilon_c$, where $p$
is the pressure, $ \epsilon$ is the energy density, $c$ is the speed
of light and $\epsilon_c$ is a constant. This EoS is maximally stiff
and predicts the most rapid increase of pressure with energy density
without violating causality.  The constant $\epsilon_c$ is the parameter
that determines the discontinuity in energy density between the low- and
high-density equations of state. Our choice of $\epsilon_c$ ensures that the energy
density is continuous and provides an upper bound on both the radius
and the maximum mass of the neutron star.

Figure~\ref{fig:maxMR} shows how the bounds on the maximum radius and mass of
the neutron star vary with our choice of the critical density $\rho_c$. It
also illustrates that the bounds provide useful constraints only when the
EoS is known up to $(2-3)\rho_0$.  In Ref. \cite{Schwenk:2010} bounds on
the radius were derived by using an EoS of neutron matter calculated up to
$\rho_0$ with specific assumptions about polytropic equations of state at
higher densities.  Our upper bounds are model independent and show that the
radius of a 1.4$M_{\rm solar}$ neutron star can be 
as large as 16 km if $\rho_c=\rho_0$. To obtain a tighter bound 
the equation of state between 1$\rho_0$ and 2$\rho_0$ is important.
The red, green, blue and black curves are predictions corresponding
to the 3n interaction strength fit to $E_{\rm sym}=30.5,~32.0,~33.7$
and $35.1$ MeV, respectively.  We also note that these bounds do not change
much for $\rho_c \gsim 4 \rho_0$ because the QMC EoS is already close to
being maximally stiff in this region. These upper bounds provide a
direct relation between the experimentally measurable nuclear symmetry
energy and the maximum possible mass and radius of neutron stars.

To summarize, we predict that the correlation between the symmetry energy
and its derivative at nuclear density is nearly independent 
of the detailed short-range 3n force once its strength is
tuned to give a particular value of $E_{\rm sym}$. 
Consequently, in our model 
one short-distance parameter $A_R$ completely determines the behavior of the EoS.
At higher density, the sensitivity to short-distance
behavior of the 3n interaction translates to an uncertainty of about $1$
km for the neutron star radius with mass $M=1.4M_{\rm solar}$.
The uncertainty at high density due to a poorly constrained symmetry
energy is larger, $\simeq 3$ km. Within our model we
 predict that neutron star radii
are in the $10-13$ km range for nuclear symmetry energy in the range $32-34$
MeV. If nuclear experiments can determine that $E_{\rm sym} \le 32 $ MeV,
QMC predicts that $L \lsim 45$ MeV at nuclear density, and for neutron
stars it predicts $M_{\rm max} < 2.2 M_{\rm solar}$ and $R <12$ km
for a neutron star with $M=1.4 M_{\rm solar}$ . The relationship
between the symmetry energy and its density dependence is 
experimentally relevant, and its implications on the neutron star mass radius relationship
are subject to clear observational tests.

We thank Bob Wiringa, Steve Pieper, Kevin Schmidt and  Francesco Pederiva,
for useful discussions. We also thank Chuck Horowitz, Jim Lattimer,
Madappa Prakash, and Achim Schwenk for comments on an early version of
the manuscript. This work was supported by a grant from the Department
of Energy (DOE) under Contracts No. DE-FC02-07ER41457 (UNEDF SciDAC)
and No. DE-AC52-06NA25396 (LANL) and the DOE topical collaboration to study
"Neutrinos and nucleosynthesis in hot and dense matter."  Computer time
was made available by Los Alamos Open Supercomputing, and by the National
Energy Research Scientific Computing Center (NERSC).

%\bibliographystyle{apsrev4-1}
%\bibliography{biblio} 

\begin{thebibliography}{10}%
\makeatletter
\providecommand \@ifxundefined [1]{%
 \ifx #1\undefined \expandafter \@firstoftwo
 \else \expandafter \@secondoftwo
\fi
}%
\providecommand \@ifnum [1]{%
 \ifnum #1\expandafter \@firstoftwo
 \else \expandafter \@secondoftwo
\fi
}%
\providecommand \enquote [1]{``#1''}%
\providecommand \bibnamefont  [1]{#1}%
\providecommand \bibfnamefont [1]{#1}%
\providecommand \citenamefont [1]{#1}%
\providecommand\href[0]{\@sanitize\@href}%
\providecommand\@href[1]{\endgroup\@@startlink{#1}\endgroup\@@href}%
\providecommand\@@href[1]{#1\@@endlink}%
\providecommand \@sanitize [0]{\begingroup\catcode`\&12\catcode`\#12\relax}%
\@ifxundefined \pdfoutput {\@firstoftwo}{%
 \@ifnum{\z@=\pdfoutput}{\@firstoftwo}{\@secondoftwo}%
}{%
 \providecommand\@@startlink[1]{\leavevmode}%
 \providecommand\@@endlink[0]{}%
}{%
 \providecommand\@@startlink[1]{%
  \leavevmode
  \pdfstartlink
   attr{/Border[0 0 1 ]/H/I/C[0 1 1]}%
   user{/Subtype/Link/A<</Type/Action/S/URI/URI(#1)>>}%
  \relax
 }%
 \providecommand\@@endlink[0]{\pdfendlink}%
}%
\providecommand \url  [0]{\begingroup\@sanitize \@url }%
\providecommand \@url [1]{\endgroup\@href {#1}{\urlprefix}}%
\providecommand \urlprefix [0]{URL }%
\providecommand \Eprint[0]{\href }%
\@ifxundefined \urlstyle {%
  \providecommand \doi [1]{doi:\discretionary{}{}{}#1}%
}{%
  \providecommand \doi [0]{doi:\discretionary{}{}{}\begingroup
  \urlstyle{rm}\Url }%
}%
\providecommand \doibase [0]{http://dx.doi.org/}%
\providecommand \Doi[1]{\href{\doibase#1}}%
\providecommand \bibAnnote [3]{%
  \BibitemShut{#1}%
  \begin{quotation}\noindent
    \textsc{Key:}\ #2\\\textsc{Annotation:}\ #3%
  \end{quotation}%
}%
\providecommand \bibAnnoteFile [2]{%
  \IfFileExists{#2}{\bibAnnote {#1} {#2} {\input{#2}}}{}%
}%
\providecommand \typeout [0]{\immediate \write \m@ne }%
\providecommand \selectlanguage [0]{\@gobble}%
\providecommand \bibinfo [0]{\@secondoftwo}%
\providecommand \bibfield [0]{\@secondoftwo}%
\providecommand \translation [1]{[#1]}%
\providecommand \BibitemOpen[0]{}%
\providecommand \bibitemStop [0]{}%
\providecommand \bibitemNoStop [0]{.\EOS\space}%
\providecommand \EOS [0]{\spacefactor3000\relax}%
\providecommand \BibitemShut [1]{\csname bibitem#1\endcsname}%
%</preamble>
\bibitem{Ransom:2010}%
  \BibitemOpen
  \bibfield{author}{%
  \bibinfo {author} {\bibfnamefont{P.~B.}\ \bibnamefont{Demorest}}, \bibinfo
  {author} {\bibfnamefont{T.}~\bibnamefont{Pennucci}}, \bibinfo {author}
  {\bibfnamefont{S.~M.}\ \bibnamefont{Ransom}}, \bibinfo {author}
  {\bibfnamefont{M.~S.~E.}\ \bibnamefont{Roberts}},\ and\ \bibinfo {author}
  {\bibfnamefont{J.~W.~T.}\ \bibnamefont{Hessels}},\ }%
  \bibfield{journal}{%
  \Doi{10.1038/nature09466}{\bibinfo {journal} {Nature (London)}}\ }%
  \textbf{\bibinfo {volume} {467}},\ \bibinfo {pages} {1081} (\bibinfo {year}
  {2010})%
  \bibAnnoteFile{NoStop}{Ransom:2010}%
\bibitem{Webb:2007}%
  \BibitemOpen
  \bibfield{author}{%
  \bibinfo {author} {\bibfnamefont{N.~A.}\ \bibnamefont{{Webb}}}\ and\ \bibinfo
  {author} {\bibfnamefont{D.}~\bibnamefont{{Barret}}},\ }%
  \bibfield{journal}{%
  \Doi{10.1086/522877}{\bibinfo {journal} {Astrophys. J.}}\ }%
  \textbf{\bibinfo {volume} {671}},\ \bibinfo {pages} {727} (\bibinfo {year}
  {2007})%
  \bibAnnoteFile{NoStop}{Webb:2007}%
\bibitem{Guver:2010}%
  \BibitemOpen
  \bibfield{author}{%
  \bibinfo {author} {\bibfnamefont{T.}~\bibnamefont{{G{\"u}ver}}}, \bibinfo
  {author} {\bibfnamefont{P.}~\bibnamefont{{Wroblewski}}}, \bibinfo {author}
  {\bibfnamefont{L.}~\bibnamefont{{Camarota}}},\ and\ \bibinfo {author}
  {\bibfnamefont{F.}~\bibnamefont{{{\"O}zel}}},\ }%
  \bibfield{journal}{%
  \Doi{10.1088/0004-637X/719/2/1807}{\bibinfo {journal} {Astrophys. J.}}\ }%
  \textbf{\bibinfo {volume} {719}},\ \bibinfo {pages} {1807} (\bibinfo {year}
  {2010})%
  \bibAnnoteFile{NoStop}{Guver:2010}%
\bibitem{Steiner:2010}%
  \BibitemOpen
  \bibfield{author}{%
  \bibinfo {author} {\bibfnamefont{A.~W.}\ \bibnamefont{{Steiner}}}, \bibinfo
  {author} {\bibfnamefont{J.~M.}\ \bibnamefont{{Lattimer}}},\ and\ \bibinfo
  {author} {\bibfnamefont{E.~F.}\ \bibnamefont{{Brown}}},\ }%
  \bibfield{journal}{%
  \Doi{10.1088/0004-637X/722/1/33}{\bibinfo {journal} {Astrophys. J.}}\ }%
  \textbf{\bibinfo {volume} {722}},\ \bibinfo {pages} {33} (\bibinfo {year}
  {2010})%
  \bibAnnoteFile{NoStop}{Steiner:2010}%
\bibitem{Pieper:2001}%
  \BibitemOpen
  \bibfield{author}{%
  \bibinfo {author} {\bibfnamefont{S.~C.}\ \bibnamefont{{Pieper}}}, \bibinfo
  {author} {\bibfnamefont{V.~R.}\ \bibnamefont{{Pandharipande}}}, \bibinfo
  {author} {\bibfnamefont{R.~B.}\ \bibnamefont{{Wiringa}}},\ and\ \bibinfo
  {author} {\bibfnamefont{J.}~\bibnamefont{{Carlson}}},\ }%
  \bibfield{journal}{%
  \Doi{10.1103/PhysRevC.64.014001}{\bibinfo {journal} {Phys. Rev. C}}\ }%
  \textbf{\bibinfo {volume} {64}},\ \bibinfo {pages} {014001} (\bibinfo {year}
  {2001})%
  \bibAnnoteFile{NoStop}{Pieper:2001}%
\bibitem{Steiner:2005}%
  \BibitemOpen
  \bibfield{author}{%
  \bibinfo {author} {\bibfnamefont{A.~W.}\ \bibnamefont{{Steiner}}}, \bibinfo
  {author} {\bibfnamefont{M.}~\bibnamefont{{Prakash}}}, \bibinfo {author}
  {\bibfnamefont{J.~M.}\ \bibnamefont{{Lattimer}}},\ and\ \bibinfo {author}
  {\bibfnamefont{P.~J.}\ \bibnamefont{{Ellis}}},\ }%
  \bibfield{journal}{%
  \Doi{10.1016/j.physrep.2005.02.004}{\bibinfo {journal} {Phys. Rep.}}\ }%
  \textbf{\bibinfo {volume} {411}},\ \bibinfo {pages} {325} (\bibinfo {year}
  {2005})%
  \bibAnnoteFile{NoStop}{Steiner:2005}%
\bibitem{Shetty:2007}%
  \BibitemOpen
  \bibfield{author}{%
  \bibinfo {author} {\bibfnamefont{D.~V.}\ \bibnamefont{Shetty}}, \bibinfo
  {author} {\bibfnamefont{S.~J.}\ \bibnamefont{Yennello}},\ and\ \bibinfo
  {author} {\bibfnamefont{G.~A.}\ \bibnamefont{Souliotis}},\ }%
  \bibfield{journal}{%
  \Doi{10.1103/PhysRevC.76.024606}{\bibinfo {journal} {Phys. Rev. C}}\ }%
  \textbf{\bibinfo {volume} {76}},\ \bibinfo {pages} {024606} (\bibinfo {year}
  {2007})%
  \bibAnnoteFile{NoStop}{Shetty:2007}%
\bibitem{Schmidt:1999}%
  \BibitemOpen
  \bibfield{author}{%
  \bibinfo {author} {\bibfnamefont{K.~E.}\ \bibnamefont{Schmidt}}\ and\
  \bibinfo {author} {\bibfnamefont{S.}~\bibnamefont{Fantoni}},\ }%
  \bibfield{journal}{%
  \Doi{doi:10.1016/S0370-2693(98)01522-6}{\bibinfo {journal} {Phys. Lett. B}}\
  }%
  \textbf{\bibinfo {volume} {446}},\ \bibinfo {pages} {99} (\bibinfo {year}
  {1999})%
  \bibAnnoteFile{NoStop}{Schmidt:1999}%
\bibitem{Gandolfi:2007}%
  \BibitemOpen
  \bibfield{author}{%
  \bibinfo {author} {\bibfnamefont{S.}~\bibnamefont{Gandolfi}}, \bibinfo
  {author} {\bibfnamefont{F.}~\bibnamefont{Pederiva}}, \bibinfo {author}
  {\bibfnamefont{S.}~\bibnamefont{Fantoni}},\ and\ \bibinfo {author}
  {\bibfnamefont{K.~E.}\ \bibnamefont{Schmidt}},\ }%
  \bibfield{journal}{%
  \Doi{10.1103/PhysRevLett.99.022507}{\bibinfo {journal} {Phys. Rev. Lett.}}\
  }%
  \textbf{\bibinfo {volume} {99}},\ \bibinfo {pages} {022507} (\bibinfo {year}
  {2007})%
  \bibAnnoteFile{NoStop}{Gandolfi:2007}%
\bibitem{Sarsa:2003}%
  \BibitemOpen
  \bibfield{author}{%
  \bibinfo {author} {\bibfnamefont{A.}~\bibnamefont{{Sarsa}}}, \bibinfo
  {author} {\bibfnamefont{S.}~\bibnamefont{{Fantoni}}}, \bibinfo {author}
  {\bibfnamefont{K.~E.}\ \bibnamefont{{Schmidt}}},\ and\ \bibinfo {author}
  {\bibfnamefont{F.}~\bibnamefont{{Pederiva}}},\ }%
  \bibfield{journal}{%
  \Doi{10.1103/PhysRevC.68.024308}{\bibinfo {journal} {Phys. Rev. C}}\ }%
  \textbf{\bibinfo {volume} {68}},\ \bibinfo {pages} {024308} (\bibinfo {year}
  {2003})%
  \bibAnnoteFile{NoStop}{Sarsa:2003}%
\bibitem{Gandolfi:2011}%
  \BibitemOpen
  \bibfield{author}{%
  \bibinfo {author} {\bibfnamefont{S.}~\bibnamefont{Gandolfi}}, \bibinfo
  {author} {\bibfnamefont{J.}~\bibnamefont{Carlson}},\ and\ \bibinfo {author}
  {\bibfnamefont{S.~C.}\ \bibnamefont{Pieper}},\ }%
  \bibfield{journal}{%
  \Doi{10.1103/PhysRevLett.106.012501}{\bibinfo {journal} {Phys. Rev. Lett.}}\
  }%
  \textbf{\bibinfo {volume} {106}},\ \bibinfo {pages} {012501} (\bibinfo {year}
  {2011})%
  \bibAnnoteFile{NoStop}{Gandolfi:2011}%
\bibitem{Lattimer:2001}%
  \BibitemOpen
  \bibfield{author}{%
  \bibinfo {author} {\bibfnamefont{J.~M.}\ \bibnamefont{{Lattimer}}}\ and\
  \bibinfo {author} {\bibfnamefont{M.}~\bibnamefont{{Prakash}}},\ }%
  \bibfield{journal}{%
  \Doi{10.1086/319702}{\bibinfo {journal} {Astrophys. J.}}\ }%
  \textbf{\bibinfo {volume} {550}},\ \bibinfo {pages} {426} (\bibinfo {year}
  {2001})%
  \bibAnnoteFile{NoStop}{Lattimer:2001}%
\bibitem{Wiringa:1995}%
  \BibitemOpen
  \bibfield{author}{%
  \bibinfo {author} {\bibfnamefont{R.~B.}\ \bibnamefont{{Wiringa}}}, \bibinfo
  {author} {\bibfnamefont{V.~G.~J.}\ \bibnamefont{{Stoks}}},\ and\ \bibinfo
  {author} {\bibfnamefont{R.}~\bibnamefont{{Schiavilla}}},\ }%
  \bibfield{journal}{%
  \Doi{10.1103/PhysRevC.51.38}{\bibinfo {journal} {Phys. Rev. C}}\ }%
  \textbf{\bibinfo {volume} {51}},\ \bibinfo {pages} {38} (\bibinfo {year}
  {1995})%
  \bibAnnoteFile{NoStop}{Wiringa:1995}%
\bibitem{BobPrivate}%
  \BibitemOpen
  \bibfield{author}{%
  \bibinfo {author} {\bibfnamefont{R.~B.}\ \bibnamefont{{Wiringa}}}\ }%
  \bibinfo {note} {private Communicaion},\
  \url{http://www.phy.anl.gov/theory/research/av18/}%
  \bibAnnoteFile{NoStop}{BobPrivate}%
\bibitem{Schiavilla:2007}%
  \BibitemOpen
  \bibfield{author}{%
  \bibinfo {author} {\bibfnamefont{R.}~\bibnamefont{{Schiavilla}}}, \bibinfo
  {author} {\bibfnamefont{R.~B.}\ \bibnamefont{{Wiringa}}}, \bibinfo {author}
  {\bibfnamefont{S.~C.}\ \bibnamefont{{Pieper}}},\ and\ \bibinfo {author}
  {\bibfnamefont{J.}~\bibnamefont{{Carlson}}},\ }%
  \bibfield{journal}{%
  \Doi{10.1103/PhysRevLett.98.132501}{\bibinfo {journal} {Phys. Rev. Lett.}}\
  }%
  \textbf{\bibinfo {volume} {98}},\ \bibinfo {pages} {132501} (\bibinfo {year}
  {2007})%
  \bibAnnoteFile{NoStop}{Schiavilla:2007}%
\bibitem{Piasetzky:2006}%
  \BibitemOpen
  \bibfield{author}{%
  \bibinfo {author} {\bibfnamefont{E.}~\bibnamefont{{Piasetzky}}}, \bibinfo
  {author} {\bibfnamefont{M.}~\bibnamefont{{Sargsian}}}, \bibinfo {author}
  {\bibfnamefont{L.}~\bibnamefont{{Frankfurt}}}, \bibinfo {author}
  {\bibfnamefont{M.}~\bibnamefont{{Strikman}}},\ and\ \bibinfo {author}
  {\bibfnamefont{J.~W.}\ \bibnamefont{{Watson}}},\ }%
  \bibfield{journal}{%
  \Doi{10.1103/PhysRevLett.97.162504}{\bibinfo {journal} {Phys. Rev. Lett.}}\
  }%
  \textbf{\bibinfo {volume} {97}},\ \bibinfo {pages} {162504} (\bibinfo {year}
  {2006})%
  \bibAnnoteFile{NoStop}{Piasetzky:2006}%
\bibitem{Tang:2003}%
  \BibitemOpen
  \bibfield{author}{%
  \bibinfo {author} {\bibfnamefont{A.}~\bibnamefont{{Tang}}}, \bibinfo {author}
  {\bibfnamefont{J.~W.}\ \bibnamefont{{Watson}}}, \bibinfo {author}
  {\bibfnamefont{J.}~\bibnamefont{{Aclander}}}, \bibinfo {author}
  {\bibfnamefont{J.}~\bibnamefont{{Alster}}}, \bibinfo {author}
  {\bibfnamefont{G.}~\bibnamefont{{Asryan}}}, \bibinfo {author}
  {\bibfnamefont{Y.}~\bibnamefont{{Averichev}}}, \bibinfo {author}
  {\bibfnamefont{D.}~\bibnamefont{{Barton}}}, \bibinfo {author}
  {\bibfnamefont{V.}~\bibnamefont{{Baturin}}}, \bibinfo {author}
  {\bibfnamefont{N.}~\bibnamefont{{Bukhtoyarova}}}, \bibinfo {author}
  {\bibfnamefont{A.}~\bibnamefont{{Carroll}}}, \bibinfo {author}
  {\bibfnamefont{S.}~\bibnamefont{{Gushue}}}, \bibinfo {author}
  {\bibfnamefont{S.}~\bibnamefont{{Heppelmann}}}, \bibinfo {author}
  {\bibfnamefont{A.}~\bibnamefont{{Leksanov}}}, \bibinfo {author}
  {\bibfnamefont{Y.}~\bibnamefont{{Makdisi}}}, \bibinfo {author}
  {\bibfnamefont{A.}~\bibnamefont{{Malki}}}, \bibinfo {author}
  {\bibfnamefont{E.}~\bibnamefont{{Minina}}}, \bibinfo {author}
  {\bibfnamefont{I.}~\bibnamefont{{Navon}}}, \bibinfo {author}
  {\bibfnamefont{H.}~\bibnamefont{{Nicholson}}}, \bibinfo {author}
  {\bibfnamefont{A.}~\bibnamefont{{Ogawa}}}, \bibinfo {author}
  {\bibfnamefont{Y.}~\bibnamefont{{Panebratsev}}}, \bibinfo {author}
  {\bibfnamefont{E.}~\bibnamefont{{Piasetzky}}}, \bibinfo {author}
  {\bibfnamefont{A.}~\bibnamefont{{Schetkovsky}}}, \bibinfo {author}
  {\bibfnamefont{S.}~\bibnamefont{{Shimanskiy}}},\ and\ \bibinfo {author}
  {\bibfnamefont{D.}~\bibnamefont{{Zhalov}}},\ }%
  \bibfield{journal}{%
  \Doi{10.1103/PhysRevLett.90.042301}{\bibinfo {journal} {Phys. Rev. Lett.}}\
  }%
  \textbf{\bibinfo {volume} {90}},\ \bibinfo {pages} {042301} (\bibinfo {year}
  {2003})%
  \bibAnnoteFile{NoStop}{Tang:2003}%
\bibitem{Subedi:2008}%
  \BibitemOpen
  \bibfield{author}{%
  \bibinfo {author} {\bibfnamefont{R.}~\bibnamefont{{Subedi {\it et al.}}}},\
  }%
  \bibfield{journal}{%
  \Doi{10.1126/science.1156675}{\bibinfo {journal} {Science}}\ }%
  \textbf{\bibinfo {volume} {320}},\ \bibinfo {pages} {1476} (\bibinfo {year}
  {2008})%
  \bibAnnoteFile{NoStop}{Subedi:2008}%
\bibitem{Shneor:2007}%
  \BibitemOpen
  \bibfield{author}{%
  \bibinfo {author} {\bibfnamefont{R.}~\bibnamefont{{Shneor {\it et al.}}}},\
  }%
  \bibfield{journal}{%
  \Doi{10.1103/PhysRevLett.99.072501}{\bibinfo {journal} {Phys. Rev. Lett.}}\
  }%
  \textbf{\bibinfo {volume} {99}},\ \bibinfo {pages} {072501} (\bibinfo {year}
  {2007})%
  \bibAnnoteFile{NoStop}{Shneor:2007}%
\bibitem{Wiringa:2002}%
  \BibitemOpen
  \bibfield{author}{%
  \bibinfo {author} {\bibfnamefont{R.~B.}\ \bibnamefont{Wiringa}}\ and\
  \bibinfo {author} {\bibfnamefont{S.~C.}\ \bibnamefont{Pieper}},\ }%
  \bibfield{journal}{%
  \Doi{10.1103/PhysRevLett.89.182501}{\bibinfo {journal} {Phys. Rev. Lett.}}\
  }%
  \textbf{\bibinfo {volume} {89}},\ \bibinfo {pages} {182501} (\bibinfo {year}
  {2002})%
  \bibAnnoteFile{NoStop}{Wiringa:2002}%
\bibitem{Pieper}%
  \BibitemOpen
  \bibfield{author}{%
  \bibinfo {author} {\bibfnamefont{S.~C.}\ \bibnamefont{{Pieper}}}\ and\
  \bibinfo {author} {\bibfnamefont{R.~B.}\ \bibnamefont{{Wiringa}}}\ }%
  \bibinfo {note} {private Communication}%
  \bibAnnoteFile{NoStop}{Pieper}%
\bibitem{Pieper:2008}%
  \BibitemOpen
  \bibfield{author}{%
  \bibinfo {author} {\bibfnamefont{S.~C.}\ \bibnamefont{Pieper}},\ }%
  \bibfield{journal}{%
  \Doi{10.1063/1.2932280}{\bibinfo {journal} {AIP Conf. Proc.}}\ }%
  \textbf{\bibinfo {volume} {1011}},\ \bibinfo {pages} {143} (\bibinfo {year}
  {2008})%
  \bibAnnoteFile{NoStop}{Pieper:2008}%
\bibitem{Akmal:1998}%
  \BibitemOpen
  \bibfield{author}{%
  \bibinfo {author} {\bibfnamefont{A.}~\bibnamefont{Akmal}}, \bibinfo {author}
  {\bibfnamefont{V.~R.}\ \bibnamefont{Pandharipande}},\ and\ \bibinfo {author}
  {\bibfnamefont{D.~G.}\ \bibnamefont{Ravenhall}},\ }%
  \bibfield{journal}{%
  \Doi{10.1103/PhysRevC.58.1804}{\bibinfo {journal} {Phys. Rev. C}}\ }%
  \textbf{\bibinfo {volume} {58}},\ \bibinfo {pages} {1804} (\bibinfo {year}
  {1998})%
  \bibAnnoteFile{NoStop}{Akmal:1998}%
\bibitem{Epelbaum:2009}%
  \BibitemOpen
  \bibfield{author}{%
  \bibinfo {author} {\bibfnamefont{E.}~\bibnamefont{{Epelbaum}}}, \bibinfo
  {author} {\bibfnamefont{H.-W.}~\bibnamefont{{Hammer}}},\ and\ \bibinfo {author}
  {\bibfnamefont{Ulf-G.}~\bibnamefont{{Mei{\ss}ner}}},\ }%
  \bibfield{journal}{%
  \Doi{10.1103/RevModPhys.81.1773}{\bibinfo {journal} {Rev. Mod. Phys.}}\ }%
  \textbf{\bibinfo {volume} {81}},\ \bibinfo {pages} {1773} (\bibinfo {year}
  {2009})%
  \bibAnnoteFile{NoStop}{Epelbaum:2009}%
\bibitem{Tsang:2009}%
  \BibitemOpen
  \bibfield{author}{%
  \bibinfo {author} {\bibfnamefont{M.~B.}\ \bibnamefont{Tsang}}, \bibinfo
  {author} {\bibfnamefont{Y.}~\bibnamefont{Zhang}}, \bibinfo {author}
  {\bibfnamefont{P.}~\bibnamefont{Danielewicz}}, \bibinfo {author}
  {\bibfnamefont{M.}~\bibnamefont{Famiano}}, \bibinfo {author}
  {\bibfnamefont{Z.}~\bibnamefont{Li}}, \bibinfo {author}
  {\bibfnamefont{W.~G.}\ \bibnamefont{Lynch}},\ and\ \bibinfo {author}
  {\bibfnamefont{A.~W.}\ \bibnamefont{Steiner}},\ }%
  \bibfield{journal}{%
  \Doi{10.1103/PhysRevLett.102.122701}{\bibinfo {journal} {Phys. Rev. Lett.}}\
  }%
  \textbf{\bibinfo {volume} {102}},\ \bibinfo {pages} {122701} (\bibinfo {year}
  {2009})%
  \bibAnnoteFile{NoStop}{Tsang:2009}%
\bibitem{Moller:1995}%
  \BibitemOpen
  \bibfield{author}{%
  \bibinfo {author} {\bibfnamefont{P.}~\bibnamefont{{M{\"o}ller}}}, \bibinfo
  {author} {\bibfnamefont{J.~R.}\ \bibnamefont{{Nix}}}, \bibinfo {author}
  {\bibfnamefont{W.~D.}\ \bibnamefont{{Myers}}},\ and\ \bibinfo {author}
  {\bibfnamefont{W.~J.}\ \bibnamefont{{Swiatecki}}},\ }%
  \bibfield{journal}{%
  \Doi{10.1006/adnd.1995.1002}{\bibinfo {journal} {At. Data Nucl. Data
  Tables}}\ }%
  \textbf{\bibinfo {volume} {59}},\ \bibinfo {pages} {185} (\bibinfo {year}
  {1995})%
  \bibAnnoteFile{NoStop}{Moller:1995}%
\bibitem{Brown:2000}%
  \BibitemOpen
  \bibfield{author}{%
  \bibinfo {author} {\bibfnamefont{B.}~\bibnamefont{Alex~Brown}},\ }%
  \bibfield{journal}{%
  \Doi{10.1103/PhysRevLett.85.5296}{\bibinfo {journal} {Phys. Rev. Lett.}}\ }%
  \textbf{\bibinfo {volume} {85}},\ \bibinfo {pages} {5296} (\bibinfo {month}
  {Dec}\ \bibinfo {year} {2000})%
  \bibAnnoteFile{NoStop}{Brown:2000}%
\bibitem{Hebeler:2010}%
  \BibitemOpen
  \bibfield{author}{%
  \bibinfo {author} {\bibfnamefont{K.}~\bibnamefont{Hebeler}}\ and\ \bibinfo
  {author} {\bibfnamefont{A.}~\bibnamefont{Schwenk}},\ }%
  \bibfield{journal}{%
  \Doi{10.1103/PhysRevC.82.014314}{\bibinfo {journal} {Phys. Rev. C}}\ }%
  \textbf{\bibinfo {volume} {82}},\ \bibinfo {pages} {014314} (\bibinfo {year}
  {2010})%
  \bibAnnoteFile{NoStop}{Hebeler:2010}%
\bibitem{Gandolfi:2009}%
  \BibitemOpen
  \bibfield{author}{%
  \bibinfo {author} {\bibfnamefont{S.}~\bibnamefont{Gandolfi}}, \bibinfo
  {author} {\bibfnamefont{A.~Yu.}\ \bibnamefont{Illarionov}}, \bibinfo {author}
  {\bibfnamefont{K.~E.}\ \bibnamefont{Schmidt}}, \bibinfo {author}
  {\bibfnamefont{F.}~\bibnamefont{Pederiva}},\ and\ \bibinfo {author}
  {\bibfnamefont{S.}~\bibnamefont{Fantoni}},\ }%
  \bibfield{journal}{%
  \Doi{10.1103/PhysRevC.79.054005}{\bibinfo {journal} {Phys. Rev. C}}\ }%
  \textbf{\bibinfo {volume} {79}},\ \bibinfo {pages} {054005} (\bibinfo {year}
  {2009})%
  \bibAnnoteFile{NoStop}{Gandolfi:2009}%
\bibitem{Lattimer:2004}%
  \BibitemOpen
  \bibfield{author}{%
  \bibinfo {author} {\bibfnamefont{J.~M.}\ \bibnamefont{{Lattimer}}}\ and\
  \bibinfo {author} {\bibfnamefont{M.}~\bibnamefont{{Prakash}}},\ }%
  \bibfield{journal}{%
  \Doi{10.1126/science.1090720}{\bibinfo {journal} {Science}}\ }%
  \textbf{\bibinfo {volume} {304}},\ \bibinfo {pages} {536} (\bibinfo {year}
  {2004})%
  \bibAnnoteFile{NoStop}{Lattimer:2004}%
\bibitem{Gandolfi:2010}%
  \BibitemOpen
  \bibfield{author}{%
  \bibinfo {author} {\bibfnamefont{S.}~\bibnamefont{Gandolfi}}, \bibinfo
  {author} {\bibfnamefont{A.~Yu.}\ \bibnamefont{Illarionov}}, \bibinfo {author}
  {\bibfnamefont{S.}~\bibnamefont{Fantoni}}, \bibinfo {author}
  {\bibfnamefont{J.~C.}~\bibnamefont{Miller}}, \bibinfo {author}
  {\bibfnamefont{F.}~\bibnamefont{Pederiva}},\ and\ \bibinfo {author}
  {\bibfnamefont{K.~E.}~\bibnamefont{Schmidt}},\ }%
  \bibfield{journal}{%
  \Doi{10.1111/j.1745-3933.2010.00829.x}{\bibinfo {journal} {Mon. Not. R.
  Astron. Soc.}}\ }%
  \textbf{\bibinfo {volume} {404}},\ \bibinfo {pages} {L35} (\bibinfo {year}
  {2010})%
  \bibAnnoteFile{NoStop}{Gandolfi:2010}%
\bibitem{BPS:1971}%
  \BibitemOpen
  \bibfield{author}{%
  \bibinfo {author} {\bibfnamefont{G.}~\bibnamefont{{Baym}}}, \bibinfo {author}
  {\bibfnamefont{C.}~\bibnamefont{{Pethick}}},\ and\ \bibinfo {author}
  {\bibfnamefont{P.}~\bibnamefont{{Sutherland}}},\ }%
  \bibfield{journal}{%
  \Doi{10.1086/151216}{\bibinfo {journal} {Astrophys. J.}}\ }%
  \textbf{\bibinfo {volume} {170}},\ \bibinfo {pages} {299} (\bibinfo {year}
  {1971})%
  \bibAnnoteFile{NoStop}{BPS:1971}%
\bibitem{NegeleVautherin:1973}%
  \BibitemOpen
  \bibfield{author}{%
  \bibinfo {author} {\bibfnamefont{J.~W.}\ \bibnamefont{{Negele}}}\ and\
  \bibinfo {author} {\bibfnamefont{D.}~\bibnamefont{{Vautherin}}},\ }%
  \bibfield{journal}{%
  \Doi{10.1016/0375-9474(73)90349-7}{\bibinfo {journal} {Nucl. Phys. A}}\ }%
  \textbf{\bibinfo {volume} {207}},\ \bibinfo {pages} {298} (\bibinfo {year}
  {1973})%
  \bibAnnoteFile{NoStop}{NegeleVautherin:1973}%
\bibitem{RhoadesRuffini:1974}%
  \BibitemOpen
  \bibfield{author}{%
  \bibinfo {author} {\bibfnamefont{C.~E.}\ \bibnamefont{{Rhoades}}}\ and\
  \bibinfo {author} {\bibfnamefont{R.}~\bibnamefont{{Ruffini}}},\ }%
  \bibfield{journal}{%
  \Doi{10.1103/PhysRevLett.32.324}{\bibinfo {journal} {Phys. Rev. Lett.}}\ }%
  \textbf{\bibinfo {volume} {32}},\ \bibinfo {pages} {324} (\bibinfo {year}
  {1974})%
  \bibAnnoteFile{NoStop}{RhoadesRuffini:1974}%
\bibitem{Schwenk:2010}%
  \BibitemOpen
  \bibfield{author}{%
  \bibinfo {author} {\bibfnamefont{K.}~\bibnamefont{Hebeler}}, \bibinfo
  {author} {\bibfnamefont{J.~M.}\ \bibnamefont{Lattimer}}, \bibinfo {author}
  {\bibfnamefont{C.~J.}\ \bibnamefont{Pethick}},\ and\ \bibinfo {author}
  {\bibfnamefont{A.}~\bibnamefont{{Schwenk}}},\ }%
  \bibfield{journal}{%
  \Doi{10.1103/PhysRevLett.105.161102}{\bibinfo {journal} {Phys. Rev. Lett.}}\
  }%
  \textbf{\bibinfo {volume} {105}},\ \bibinfo {pages} {161102} (\bibinfo {year}
  {2010})%
  \bibAnnoteFile{NoStop}{Schwenk:2010}%
\end{thebibliography}

%Merlin.mbs v4.21 2009-07-09.
%

\end{document}